\documentclass[twoside]{ilcws08}
\usepackage[latin1]{inputenc}
\usepackage[dvips]{graphicx,epsfig,color}
\usepackage{wrapfig,rotating}
\usepackage{amssymb,amsmath,array}
\usepackage{subfigure}

\begin{document}
\title{Study of Solid State Photon Detectors Read Out of Scintillator Tiles}
\author{A. Calcaterra$^1$, R. de Sangro$^1$\cite{url}, G. Finocchiaro$^1$, 
E. Kuznetsova$^2$, P. Patteri$^1$ and M. Piccolo$^1$ 
\vspace{.3cm}\\
1- INFN, Laboratori Nazionali di Frascati, Frascati, Italy
\vspace{.1cm}\\
2- INFN, Sezione di Roma 1, Roma, Italy}

\maketitle

\begin{abstract}
We present preliminary results on efficiency and light collection
uniformity read out performances of different
assemblies of scintillator tiles, coupled with solid state photon detectors of
different make. Our test beam data suggest that the use of 2\,mm tiles without
wavelength shifting fibers may be possible in an ILC hadron calorimeter. 
\end{abstract}

\section{Introduction}
The present work is part of an ongoing R\&D activity on ILC calorimetry 
at INFN Laboratori Nazionali di Frascati motivated by several issues pointed out by
previous experiences, particularly the CALICE hadron calorimeter
which is discussed elsewhere\,\cite{caliceProc} in these proceedings.
\par
The CALICE prototype\,\cite{caliceColl} is made of $\approx
8000$, 5\,mm thick, scintillator tiles each with a wavelength shifter
(WS) fiber delivering green-shifted scintillation light to a MEPhI/Pulsar silicon
photo-multiplier (SiPM)\,\cite{mephisipm} attached to one of its ends. The
WS fiber is positioned in an arc shaped groove individually milled in each
tile, and its function is to shift the wavelength of the
scintillation light toward the green where the quantum efficiency of the
MEPhI SiPM is highest, and to improve the light collection efficiency.
\par
One question investigated here regards the use of WS fibers, which may become an
engineering and cost challenge for the construction of a full scale, several
million tiles, ILC calorimeter. 
This prompted the study of alternative readout solutions,
without the use of WS fibers, to simplify the detector construction. 
Several companies\,\cite{HAM} are now building solid state photon devices
similar to the original MEPhI SiPM (i.e. Hamamatsu
MPPC's, ITC-IRST SRD's, SenSL SPM's), which are also sensitive 
to blue light and make the WS fiber even less necessary.
As the cost of an ILC detector scales up with size, a second aspect of this
work relates to the
performances of thinner scintillator tiles, which could reduce the total detector volume.

\section{Test Set Up and Calibrations}
We exposed to a beam of $\approx 500$\,MeV electrons, of about (10$\times$5)\,mm$^2$ RMS 
transverse size, produced at the Frascati
Beam Test Facility(BTF)\,\cite{BTF}, seven different assemblies of
scintillator tiles coupled to different silicon photon
detectors (PD). 
\begin{table}
\begin{footnotesize}
\centerline{\begin{tabular}{|clcrrcccc|}\hline
Config. & Scint. Type & Tile Thick. & PD (tot. area) & \# of pxl & pxl size & V$_{brk}$(V) & V$_{bias}$(V)& G(10$^6$)\\\hline
1& BC400 & 5\,mm & Hamam. (1\,mm$^2$) & 400 & 50$\mu$m & 68.1 & 2.6 & 1.3 \\
2& BC400 & 5\,mm & Hamam. (1\,mm$^2$) & 1,600 & 25$\mu$m & 69.6 & 2.1 & 0.5\\
3& Vladimir & 5\,mm & Hamam. (1\,mm$^2$) & 400 & 50$\mu$m & 68.9 & 1.8 & 1.1\\
4& EJ212 & 2\,mm & Hamam. (1\,mm$^2$) & 400 & 50$\mu$m & 69.0 & 1.6 & 2.2 \\
5& EJ212 & 2\,mm & Hamam. (1\,mm$^2$) & 1,600 & 25$\mu$m & 68.5 & 3.3 & 0.5\\
6& BC400 & 5\,mm & Hamam. (9\,mm$^2$) & 3,600 & 50$\mu$m & 67.3 & 1.1 & 0.6\\
7& Vladimir & 5\,mm & MEPhI (1\,mm$^2$) & 1,156 & 20$\mu$m & 68.4 & 4.8 & 0.5\\\hline
\end{tabular}}
\caption{Configurations under test. See text for the definitions of V$_{brk}$, V$_{bias}$ and G.}
\label{tab:listConf}
\end{footnotesize}
\end{table}
We cut and polished a total of 6 (3$\times$3)\,cm$^2$ plastic
scintillator tiles with 2 and 5\,mm thickness, and wrapped them in
aluminized mylar and black tape; the scintillators used were the BC400
from Saint Gobain and the EJ212 from Eljen Technology which have almost identical
characteristics and a similar scintillator made in Vladimir (Russia).
We also studied, as a reference, a CALICE tile made with the Vladimir scintillator and a 1\,mm diameter Kuraray Y11
wavelength shifter fiber. We used as photon detectors three Hamamatsu
MPPC types, differing in the total number of square pixels and individual
sizes: three 400, 50\,$\mu$m$^2$ pixels, 
two 1600, 25\,$\mu$m$^2$ pixels 
and one 3600, 50\,$\mu$m$^2$ pixels; 
the CALICE tile was read out with a MEPhI/Pulsar SiPM with
1156, 20\,$\mu$m$^2$ pixels.
Other configurations with the ITC-IRST and SensL
PD's are currently under study, and will be presented in a future paper. 
The characteristics of the configurations discussed in this paper are summarized in Tab.~\ref{tab:listConf}.
\begin{wrapfigure}{rl}{0.45\columnwidth}
\centerline{\includegraphics[width=0.45\columnwidth]{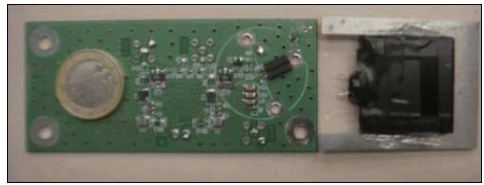}}
\caption{A tile-PD assembly coupled to the amplifier board.}
\label{fig:sipmampli}
\end{wrapfigure}
\par
The PD was attached with optical glue directly to the middle of one {\it
side} of the tile in all the different configurations
except for the (3$\times$3\,mm$^2$), 3600 pixel Hamamatsu MPPC which was
instead glued to the center 
of a {\it face}, and the CALICE tile, which has the MEPhI/Pulsar
SiPM mechanically coupled, without glue or optical grease, to one of the ends of the WS fiber.\par
The PD's were read out with a low noise $\times$10 amplifier built in Frascati, an INFN-Pisa design
based on the GALI-5 chip,  which was 
connected to the tile-PD assembly as shown in Fig.~\ref{fig:sipmampli}.
The seven tiles were then mounted in a test box where they were aligned
to each other and kept in a fixed position. The box was cabled to
provide low voltage power to the amplifiers, voltage bias to the PD's and
to extract the amplified signals. It was also equipped with a
temperature probe to monitor the operating temperature with a 
typical resolution of $\simeq 0.2^\circ C$.\par
\begin{wrapfigure}{lr}{0.45\columnwidth}
\centerline{\includegraphics[width=0.45\columnwidth]{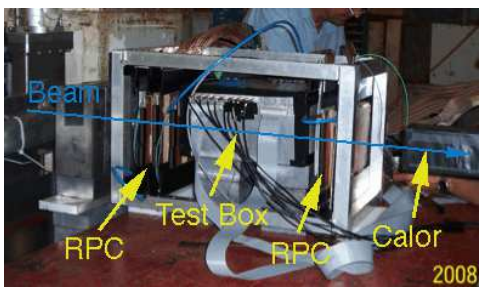}}
\caption{Test beam set up.}
\label{fig:beamsetup}
\end{wrapfigure}
Over the whole period of data taking the temperature
in the experimental hall was regulated to $23.4\pm 0.5^\circ$C, and the air
temperature inside the box has been kept constant at $26.2\pm0.25^\circ$C using a
Peltier cell in thermal contact with the box to extract some of the heat produced
by the amplifiers (about 1\,W/channel).
\par
To measure the impact point of the beam on the tile we used an external tracker
including 5 glass RPC\,\cite{rpc}, 3 of which were placed in front and 2 behind the test
box on the beam line. The RPC were equipped with orthogonal planes of strips
8\,mm wide, digitally read out, providing X-Y measurement in each
plane with a point resolution of $\simeq$2.3\,mm.
As the beam at the Frascati BTF can provide a tunable number of particles
per pulse (1-1000)\,\cite{BTF}, the test setup included a lead glass
calorimeter module to measure the beam total energy on a pulse by pulse 
basis and allowing the selection of events containing any number of 
electrons (0,1,...n). 
\begin{figure}
\centerline{\includegraphics[width=1.0\columnwidth]{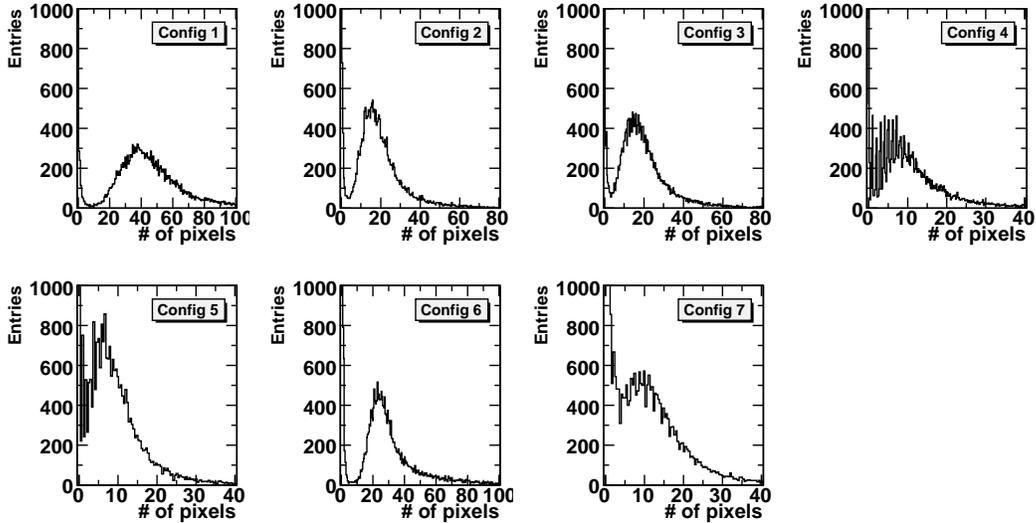}}
\caption{Plots of the signal (number of pixels) collected by the 
configurations listed in Tab.\,\ref{tab:listConf}, which also lists the
values of the set $V_{bias}$.}\label{fig:sipmPixel}
\end{figure}
A picture of the set up in the beam line is shown in Fig.~\ref{fig:beamsetup}.
\par
\begin{wrapfigure}{r}{0.45\columnwidth}
\centerline{\includegraphics[width=0.45\columnwidth]{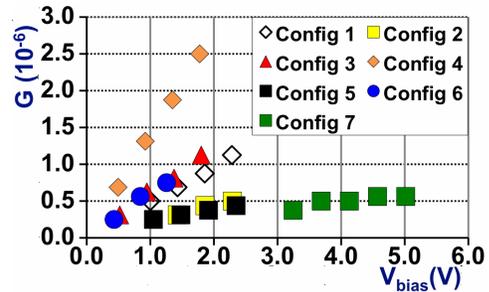}}
\caption{Plot of the gain ($Q_{1pxl}/e$) as a function of $V_{bias}$.}\label{fig:gvsbias}
\end{wrapfigure}
The response of the tiles under test to 1 MIP, in terms of the number of
fired pixels, is given in Fig.~\ref{fig:sipmPixel}.
To obtain these distributions, the
integrated charge corresponding to one pixel (PD gain) must be measured. 
Such measurement can be obtained from the
charge distributions of background events selected
requiring a pedestal reading from the lead-glass calorimeter (0 MIPs). 
In these distributions a large pedestal peak is accompanied by smaller
but distinguishable charge peaks corresponding to 1 or occasionally 2 pixels, 
fired due to the thermal noise of the devices (typical singles rate being a few 100\,kHz to 1\,MHz); 
the gain is measured by the distance of these peaks from the pedestal. 
We repeated this measurement for different bias voltages and
found the breakdown voltage of each device, defined as the voltage corresponding to zero gain. 
In Fig.~\ref{fig:gvsbias} we show the plot of the gain as
a function $V_{bias}=V-V_{brk}$; as expected all the PD's show good
linearity. 
All the data shown in this paper were taken with the same values for $V_{bias}$ listed
in Tab.~\ref{tab:listConf}, which also shows the corresponding gain expressed
in number of electrons ($Q_{1pxl}/e$).
\section{Results and Conclusions}
\begin{figure}
\centerline{\includegraphics[width=1.0\columnwidth]{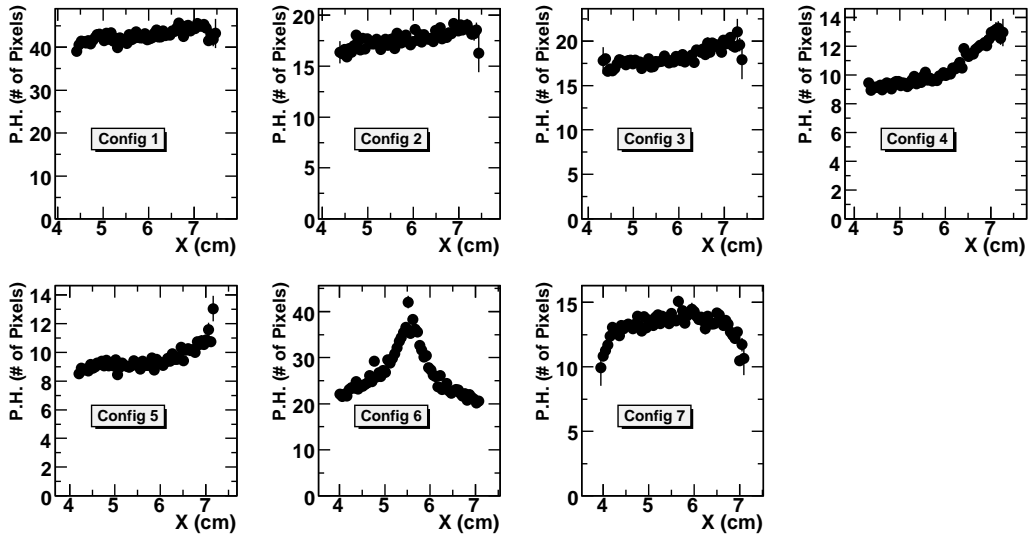}}
\caption{Profile histograms of the signal (number of pixels) as a function of the MIP
impact point of the tile, collected by the seven different
test configurations of Tab.~\ref{tab:listConf}.}\label{fig:pixelprof}
\end{figure}

\begin{figure}
\centerline{\includegraphics[width=1.0\columnwidth]{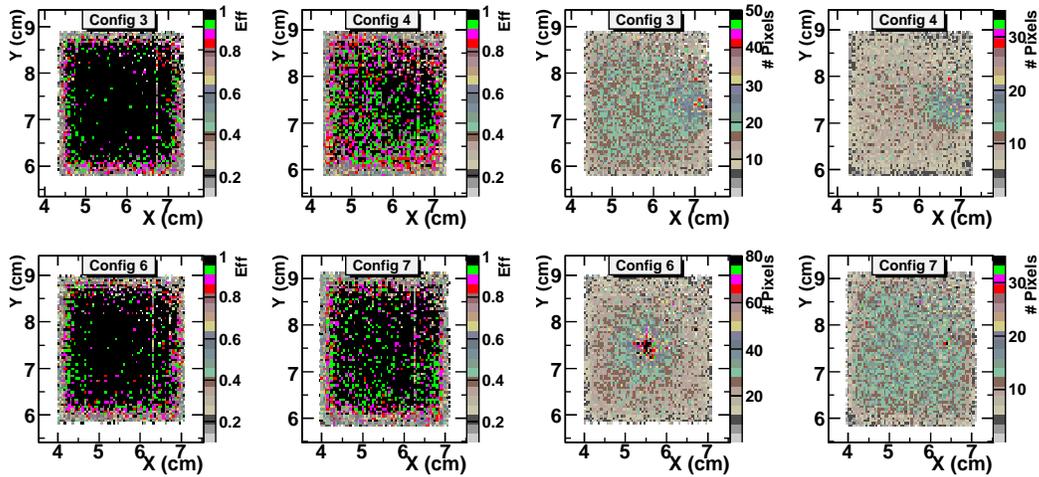}}
\caption{The leftmost four plots show the efficiency as a function of the X,Y
coordinates of the impact point on the tile for
configurations number 3,4,6,7 of Tab.~\ref{tab:listConf}. The rightmost four
plots show the pulse height in number of pixels as a function of X,Y.}\label{fig:effi2D}
\end{figure}

We have collected several million events containing exactly
1 MIP, and studied the performances of the various devices as a 
function of the beam particle impact point on the tiles. 
In Fig.~\ref{fig:pixelprof} we show the amplitude of the PD signal (number of
pixels), which is proportional to the amount of scintillation
light collected, as a function of one coordinate, whereas in the four
leftmost plots of 
Fig.~\ref{fig:effi2D} we show the efficiency in two dimensions. The
efficiency is defined by the number of times a signal above 2 pixels is
observed in the PD over the number of times a particle has crossed the
corresponding tile;
the chosen threshold corresponds to about 1/8 to 1/4 of a MIP signal,
depending on the configuration.  
\par 
From Fig.~\ref{fig:pixelprof}, we can see that all the tiles read out without using the
fiber show a somewhat higher non-uniformity in light collection, but
without loss of efficiency. 
Is evident that the
signal amplitude is maximal when the impinging particle is closest to
the PD, and decreases with distance. 
This effect is also visible for the CALICE tile, 
where the light is collected by the fiber, and where 
a degradation of light collection is observed near the edges . 
We estimate a rather high non-unifomity of $\simeq 35\%$
when the PD is attached to the face of the tile (config. 6), while config. 3
and 4 are more uniform ($\simeq 15\%$ and $\simeq 20\%$ respectively). 
The latter two values are small compared to the intrinsic fluctuations of a
MIP energy deposit, therefore their effect on an energy measurement should also be small.
In the four rightmost plots of
Fig.~\ref{fig:effi2D} we show the X,Y distribution of the pulse
height for configurations number 3,4,6 and 7.
As one can see, the largest response variation is restricted to a
small region near the position of the PD, being quite uniform elsewhere. 
This means that only a small fraction of all particles crossing a
tile is affected by this non-uniformity.
The difference in efficiency between the 5\,mm and the 2\,mm tile is
evident from the two upper leftmost plots of Fig.~\ref{fig:effi2D} (config. 3 vs 4); 
nevertheless, an efficiency greater than $\simeq 90-95\%$ over a large portion of
the tile is observed even with the thinner scintillator. 
\par
These preliminary results suggest that direct read out of
scintillator tiles with silicon PD for an ILC hadron 
calorimeter application is possible even using very thin tiles, and prompt
for detailed Monte Carlo studies to estimate their performances in a detector. 
\par
This R\&D program will continue in the future with the study of more PD and
configurations. The tunability of the
number of particles in the beam, peculiar to the Frascati BTF, will
also allow studies of the dynamic range of each tile-PD read out
configuration. Future studies will include measurements of the timing
performances of these read out schemes.

\section{Acknowledgments}
We would like to thank Dr. V. Rusinov, ITEP Moscow, for many useful suggestions on the
use of the CALICE tile assembly; L. Daniello and G. Mazzenga for their
technical skills and G. Papalino for the assembly of the test box and
associated electronics. 
\begin{footnotesize}

\end{footnotesize}

\end{document}